\documentclass[aps, prl, reprint, amsmath, amssymb, superscriptaddress,nolongbibliography]{revtex4-2}

\usepackage{graphicx}
\usepackage{dcolumn}
\usepackage{bm}
\usepackage{hyperref}
\usepackage{color}
\usepackage{mathrsfs}

\begin{document}

\preprint{APS/123-QED}

\title{Durotaxis in viscoelastic fluids}

\author{Vaseem A. Shaik}
\affiliation{Department of Mechanical Engineering, 
University of British Columbia, Vancouver, BC, V6T 1Z4, Canada}
\affiliation{Engineering Sciences and Applied Mathematics, Northwestern University, Evanston, IL 60208, USA}
\author{Jiahao Gong}
\affiliation{Department of Mathematics,
University of British Columbia, Vancouver, BC, V6T 1Z2, Canada}
\author{Gwynn J. Elfring}
 \email{gelfring@mech.ubc.ca}
\affiliation{Department of Mechanical Engineering, 
University of British Columbia, Vancouver, BC, V6T 1Z4, Canada}
\affiliation{Department of Mathematics,
University of British Columbia, Vancouver, BC, V6T 1Z2, Canada}

\date{\today}

\begin{abstract}
Organisms often swim through fluids that are spatially inhomogeneous. If the fluids are polymeric, gradients in polymer concentration may lead to gradients in both fluid viscosity and elasticity. In this letter, we present theoretical results for the dynamics of active particles, biological or otherwise, swimming through spatially inhomogeneous viscoelastic fluids. We model the active particles using the squirmer model, and show that spatial variations in fluid relaxation time lead to a novel mechanism for reorientation and taxis in viscoelastic fluids, which we refer to as a form of durotaxis in fluids.
\end{abstract}

\maketitle

Microorganisms and other active particles often swim through inhomogeneous environments prevalent in nature, such as gradients in chemical concentration, light, nutrients, temperature or salinity. The particles often respond to these inhomogeneities by exhibiting directed motion (or taxis) along gradients. Well known types of taxis include chemotaxis in chemical or nutrient gradients \cite{Berg1972, Berg2004}, phototaxis in light gradients \cite{Jekely2009}, and rheotaxis in fluid velocity gradients \cite{Bretherton1961,Marcos2012,Kantsler2014,Palacci2015,Michalec2017}. A thorough understanding of taxis in inhomogeneous environments can even be exploited to sort or organize the particle suspensions by controlling the specific inhomogeneities the particles encounter \cite{Arlt2018, Frangipane2018, Shaik2023}.

Recent research has explored taxis due to inhomogeneities in the mechanical properties of fluid environments, such as viscosity and density. For instance, taxis in fluid viscosity gradients, known as viscotaxis, was found to be exhibited by organisms like the bacteria \textit{Leptospira} \cite{Kaiser1975, Petrino1978, Takabe2017}, \textit{Spiroplasma} \cite{Daniels1980}, \textit{E. coli} \cite{Sherman1982}, and the green alga \textit{C. reinhardtii} \cite{Sherman1982, Stehnach2021, Coppola2021}. Viscosity gradients in the mucus layer of the intestine are believed to determine spatial organization of the intestinal bacteria \cite{Swidsinski2007}. Algae were even observed to scatter like light when interacting with sharp viscosity gradients \cite{Coppola2021}. These experimental observations were largely captured by theoretical efforts using different model active particles. Using a simple model, particles acted on by fixed thrust forces were shown to display positive viscotaxis \cite{Liebchen2018}, conversely models that include the interaction of the propulsion mechanism with viscosity gradients tend to lead to negative viscotaxis \cite{Datt2019, Shaik2021, Gong2023}. More generally, details about the geometry and propulsion strategy are important factors in determining the direction of taxis \cite{Gong2024} and energetic efficiency of motion in viscosity gradients \cite{Gong2024b}. Recently, a mechanism for taxis in fluid density gradients (or \textit{densitaxis}) was also discovered. In that work, swimmers were found to swim up, down or even normal to the density gradients depending on the mode of propulsion \cite{Shaik2024}, and as a result, density gradients may aid or hinder the diel vertical migration of organisms in the ocean.

In previous observations of viscotaxis, viscosity variations were often a result of spatial gradients in polymer concentration \cite{Kaiser1975, Petrino1978, Daniels1980, Sherman1982, Takabe2017, Swidsinski2007, Stehnach2021, Coppola2021}. However, changes in polymer concentration not only affect the fluid viscosity but can also lead to changes in fluid memory. The increase in viscosity with polymer concentration is well understood, but a similar variation is also observed for the relaxation time (or elastic modulus) \cite{Baumgartel1996, Liu2009,Giudice2017,Kim2019,Soetrisno2023}. Here we will show that spatial variations in fluid relaxation time also lead to reorientation and taxis in viscoelastic fluids, which we refer to as a form of \textit{durotaxis} in fluids. 

Durotaxis is well known in the context of cells crawling on extracellular substrates with stiffness gradients. Since its initial observation in fibroblasts \cite{Lo2000}, durotaxis has been identified in various cell types and even in cell collectives, where individual cells may not exhibit durotactic behavior \cite{Sunyer2020,Shellard2021,Espina2022}. Given the prevalence of mechanical inhomogeneities in animal bodies, durotaxis is speculated to play a role in immune responses and disease progression. However, the durotaxis of swimming cells in viscoelastic fluids with spatial variations of elastic moduli has not been previously reported. Research on swimming in viscoelastic fluids has largely been confined to homogeneous fluids with constant viscosity and relaxation time \cite{Elfring2015, Patteson2016, Li2021, Arratia2022, Spagnolie2023}. This extensive body of work has revealed how particle speed, power expenditure, efficiency, and swimming gait are affected by swimming in such homogeneous viscoelastic environments. But inhomogeneities in fluid properties can arise due to variations in polymer concentration and this will necessarily change swimmer dynamics due to hydrodynamic interactions with the surrounding fluid. In this letter, we will show that by considering spatial variations in fluid polymer concentration, we not only capture novel durotactic dynamics but also recover viscotaxis in inelastic fluids and locomotion in homogeneous viscoelastic fluids as special cases.

\begin{figure}
    \centering
    \includegraphics[width=0.75\linewidth]{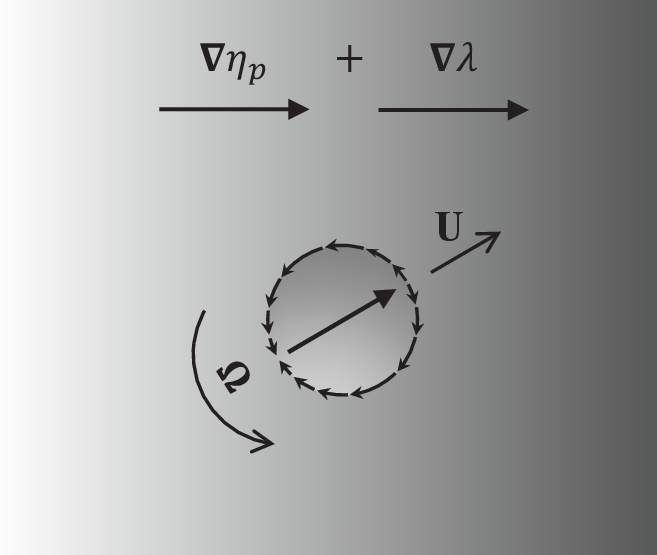}
    \caption{Schematic of an active particle swimming through gradients in the polymeric contribution to viscosity $\left( \nabla \eta_p \right)$ and relaxation time $\left( \nabla \lambda \right)$.}
    \label{fig:fig1}
\end{figure}

We consider here an active particle swimming through an inhomogeneous non-Newtonian fluid (see Fig.~\ref{fig:fig1} for a schematic). The inhomogeneity refers to spatial variations in polymer concentration within the fluid $c \left( \bf{x} \right)$, which in turn manifest as the variations in zero shear-rate polymer viscosity $\eta_p \left( \bf{x} \right)$ and relaxation time $\lambda \left( \bf{x} \right)$. These variations usually occur over length scales much larger than the particle, hence over the scale of small-sized particles considered here, we assume that the viscosity and relaxation time vary linearly
\begin{equation}
    \frac{\nabla \eta_p}{\eta_{p \infty}} = \frac{1}{L_{\eta}} {\bf{d}}, \thickspace \frac{\nabla \lambda}{\lambda_{\infty}} = \frac{1}{L_{\lambda}} {\bf{d}}.
\end{equation}
Here $\bf{d}$ is the gradient direction, $\eta_{p \infty}$ and $\lambda_{\infty}$ are the reference values chosen near the particle, while $L_{\eta}$ and $L_{\lambda}$ are the length scales over which viscosity and relaxation time vary. These length scales are related since they are both ultimately governed by the length scale over which polymer concentration is varying, however, for complete generality (and clarity) we treat them as distinct.

Neglecting fluid inertia, the incompressible flow induced by the particle satisfies the continuity equation and the Cauchy's equations of motion
\begin{align}
    \nabla \cdot {\bf{u}} = 0,\\
    \nabla \cdot \bm{\sigma} = \bf{0},
\end{align}
where $\bf{u}$ is the velocity field and $\bm{\sigma}$ is the stress tensor. The stress can be written as $\bm{\sigma} = -p {\bf{I}} + \eta_s \bm{\dot{\gamma}} + \bm{\tau}_p$, where $p$ is the pressure, $\eta_s$ is the solvent viscosity, $\bm{\dot{\gamma}} = \nabla {\bf{u}} + \left( \nabla {\bf{u}} \right)^\top$ is twice the strain-rate tensor, and $\bm{\tau}_p$ is the polymeric stress. We use the Giesekus model to describe the polymer contribution to the stress \cite{Bird1987}
\begin{equation}
    \bm{\tau}_p + \lambda \left( \bf{x} \right)  \overset{\triangledown}{\bm{\tau}}_p + \alpha \frac{\lambda \left( \bf{x} \right)}{\eta_p \left( \bf{x} \right)} \bm{\tau}_p \cdot \bm{\tau}_p = \eta_p \left( \bf{x} \right) \bm{\dot{\gamma}},
\end{equation}
where $\overset{\triangledown} {}$ denotes the upper-convected time derivative. The mobility factor $\alpha$ that takes values between $0$ and $1/2$ is related to the anisotropic Brownian motion or the anisotropic hydrodynamic drag acting on the individual polymer molecules. The Giesekus model can be derived from molecular concepts \cite{Giesekus1982}, and has also been shown to capture experimental observations \cite{DAvino2008, Snijkers2009}. This model reduces to the Oldroyd-B model when $\alpha = 0$, and to the upper-convected Maxwell model when $\alpha = 0$ and $\eta_s = 0$.

For boundary conditions, we assume the velocity field decays to zero far from the particle
\begin{equation}
    {\bf{u}} \to \bm{0} \,\, {\rm{as}} \,\, r = \left| \bf{r} \right| \to \infty,
\end{equation}
where ${\bf r} = {\bf x} - {\bf x}_c$ and ${\bf x}_c$ denotes the position of center of the particle. The active particle swims with an unknown translational velocity $\bf{U}$ and angular velocity $\bm{\Omega}$ due to its activity manifesting as some prescribed velocity on its surface ${\bf u}^s$. The boundary condition on the surface of the particle ($\partial \cal B$) is therefore
\begin{equation}
    {\bf u} \left( {\bf x} \in \partial \cal B \right) = {\bf U} + \bm{\Omega} \times {\bf r} + {\bf u}^s.
\end{equation}

We use the spherical squirmer model to describe the active particle. The squirmer model is appropriate for ciliated organisms like \textit{Paramecium} and \textit{Opalina} that propel by synchronously beating the many cilia covering their surfaces. In this model, we fix the shape of swimmer as a sphere and represent activity through a prescribed effective slip velocity on the surface of squirmer \cite{Lighthill1952, Blake1971, Ishikawa2006}
\begin{equation}
    {\bf u}^s = - \sum_{n=1}^{\infty} \frac{2}{n \left( n +1 \right)} B_n P_n' \left( {\bf p} \cdot {\bf n} \right) {\bf p} \cdot \left( {\bf I} - {\bf nn} \right).
\end{equation}
Here the coefficients $B_n$ are called the squirming modes, ${\bf p}$ is the particle orientation, ${\bf n}$ is an outward pointing unit normal to the surface of squirmer, $P_n$ is the Legendre polynomial of degree $n$, and $P_n'\left(x\right) = \frac{d}{dx}P_n\left( x \right)$. In homogeneous Newtonian fluids, the $B_1$ mode solely determines the swim speed, and here we assume $B_1>0$, while the $B_2$ mode determines the far-field force-dipole representation of the particle \cite{Nasouri2018}. The sign of $B_2$ can be used to distinguish among three types of swimmers. Pullers like \textit{C. reinhardtii} with $B_2 > 0$ swim by generating impetus in the front, pushers like \textit{E. coli} with $B_2 < 0$ propel by exerting impetus in the back, while neutral swimmers like \textit{Volvox carteri} with $B_2 = 0$ generate impetus symmetrically in both front and back, hence their flow is higher order. Here we only consider the first two squirming modes to analyze the locomotion in inhomogeneous non-Newtonian fluids, as has been a common practice in the literature on the swimming in homogeneous non-Newtonian fluids \cite{Zhu2011, Zhu2012, Li2014a, Yazdi2014, Yazdi2015, Yazdi2017} or at finite inertia \cite{Wang2012, Khair2014, Chisholm2016}.

We also neglect particle inertia, so the net force and torque acting on the particle vanish,
\begin{equation}
    {\bf F} + {\bf F}_{\rm{ext}} = \bm{0}, \thickspace {\bf L} + {\bf L}_{\rm{ext}} = \bm{0}.
\end{equation}
The hydrodynamic force and torque on the particle due to the flow it induces are
\begin{equation}
    {\bf F} = \int_{\partial \cal B} {{\bf n} \cdot \bm{\sigma}} \,\, dS, \thickspace {\bf L} = \int_{\partial \cal B} {{\bf r} \times \left({\bf n} \cdot \bm{\sigma}\right) } \,\, dS,
\end{equation}
while external forces and torques, due to buoyancy or bottom-heavyness for instance, are denoted ${\bf F}_{\rm{ext}}$, and ${\bf L}_{\rm{ext}}$.

The velocity of an active particle in a non-Newtonian fluid (homogeneous or otherwise) has been derived via the reciprocal theorem as \cite{Elfring2017}
\begin{equation}
    {\bm{\mathsf{U}}} = {{\bm{\mathsf{{{\hat R}}_{FU}^{-1}}}}} \cdot \left( {\bm{\mathsf{F}}}_{\rm{ext}} + {\bm{\mathsf{F}}}_s + {\bm{\mathsf{F}}}_{NN} \right),
\label{eqn:velocities}
\end{equation}
where ${\bm{\mathsf{U}}} = \left[{\bf U} \,\, \bm{\Omega} \right]^T$ is a six-dimensional vector containing both translational and rotational velocities. Similarly ${\bm{\mathsf{F}}} = \left[ {\bf F} \,\, {\bf L} \right]^T$ contains both force and torque, while ${{\bm{\mathsf{{{\hat R}}_{FU}}}}}$ is a $\left( 6 \times 6 \right)$ resistance tensor	 associated with a passive spherical particle in a homogeneous Newtonian fluid of viscosity $\eta_s + \eta_{p \infty}$.

The propulsive force and torque generated by the particle, in a homogeneous Newtonian fluid of the same viscosity, is 
\begin{equation}
    {\bm{\mathsf{F}}}_s = \int_{\partial \cal B} {{\bf u}^s \cdot \left( {\bf n} \cdot {\bm{\mathsf{{\hat T}_U}}} \right)} \,\, dS.
\end{equation}
All effects of inhomogeneity and the viscoelasticity of the suspending fluid are captured through the additional force and torque
\begin{equation}
    {\bm{\mathsf{F}}}_{NN} = - \int_{\cal V}{\bm{\tau}_{NN} : {\bm{\mathsf{{\hat{{E}}}_U}}} } \,\, dV,
\end{equation}
where $\bm{\tau}_{NN} = \bm{\tau}_p - \eta_{p \infty} \bm{\dot{\gamma}}$ and $\cal V$ denotes the entire fluid volume. The tensors ${\bm{\mathsf{{\hat T}_U}}}$ and ${\bm{\mathsf{{\hat{{E}}}_U}}}$ are linear operators that give the stress $\boldsymbol{\hat{\sigma}}={\bm{\mathsf{{\hat T}_U}}} \cdot {\bm{\hat{\mathsf{U}}}}$ and twice the strain rate $\boldsymbol{\hat{\dot{\gamma}}} = 2{\bm{\mathsf{{\hat{{E}}}_U}}} \cdot {\bm{\hat{\mathsf{U}}}}$ for the flow induced by a passive sphere moving with velocity ${\bm{\hat{\mathsf{U}}}}$ in a homogeneous Newtonian fluid of viscosity $\eta_s + \eta_{p \infty}$.

Using the particle radius, $a$, and the particle speed in a homogeneous Newtonian fluid, $U_N = 2B_1/3$, as characteristic scales, we find that the particle velocity in inhomogeneous non-Newtonian fluids is governed by three dimensionless numbers: the relative viscosity variations across the particle, $\varepsilon_{\eta} = a/L_{\eta}$; the relative relaxation time changes across the particle, $\varepsilon_{\lambda} = a/L_{\lambda}$; and the Deborah number, $De = \lambda_{\infty} U_N/a$, measuring the ratio of relaxation time to the time scale of swimming. We also introduce the partition parameter $\beta_\infty = \eta_{p \infty}/(\eta_s+\eta_{p \infty})$, which delineates the fraction of the viscosity due to addition of polymers in a homogeneous viscoelastic fluid. Since the viscosity and relaxation time vary over large length scales compared to particle, $\varepsilon_{\eta} \ll 1$, and $\varepsilon_{\lambda} \ll 1$. The relaxation time for biofilms, mucus and other commonly used fluids in experiments ranges from $0.4 - 10^3\,\text{s}$ \cite{Litt1976, Shaw2004, Hohne2009, Lai2009, Shen2011, Arratia2022}, and the time scale of swimming ranges from $0.1-1\,\text{s}$, hence Deborah numbers can range from $0.4 - 10^4$. Here we focus on small Deborah numbers, corresponding to swimming in weakly non-Newtonian fluids and calculate the leading order changes due to spatial variation in fluid elasticity.

Given that $De$, $\varepsilon_{\lambda}$, and $\varepsilon_{\eta}$ are all much less than 1, we expand the velocity and stress fields, as well as the particle velocity, in a regular perturbation in these small parameters. Hence we write any variable $f$ as $f = f_0 + De \, f_{100} + \varepsilon_{\lambda} \, f_{010} + \varepsilon_{\eta} \, f_{001} + \cdots$. We note that the integrals in ${\bm{\mathsf{F}}}_{NN}$ at different orders of perturbation yield finite values, as opposed to diverging integrals that would indicate a non-regular (or singular) perturbation \cite{Shaik2024}.

\textit{Passive particles.}---We consider first a passive sphere $\left({\bf u}^s = \bm{0}, \,  {\bm{\mathsf{F}}}_s = \bm{0} \right)$ towed with a fixed velocity $ {\bm{\mathsf{U}}}$. The sphere experiences a net hydrodynamic force and torque in an inhomogeneous Giesekus fluid that can be written as a correction to that in a homogeneous Newtonian fluid $ {\bm{\mathsf{F}}}_0$. Specifically, ${\bm{\mathsf{F}}} =  {\bm{\mathsf{F}}}_0 +  {\bm{\mathsf{F}}}_{NN}$, where to the leading order in $De$, $\varepsilon_{\eta}$, and $\varepsilon_{\lambda}$, the correction ${\bm{\mathsf{F}}}_{NN}$ is simply a sum of corrections in homogeneous Giesekus fluid $ \left( {\bm{\mathsf{F}}}_{NN_\infty} \right)$, viscosity gradients  $ \left( {\bm{\mathsf{F}}}_{\eta} \right)$, and stiffness gradients $ \left( {\bm{\mathsf{F}}}_{\lambda} \right)$
\begin{equation}
     {\bm{\mathsf{F}}}_{NN} =  {\bm{\mathsf{F}}}_{NN_\infty} +  {\bm{\mathsf{F}}}_{\eta} +  {\bm{\mathsf{F}}}_{\lambda}.
\end{equation}

The additional force and torque in a homogeneous Giesekus (or Oldroyd-B) fluid is well characterized \cite{Leslie1961,Giesekus1963,Thomas1964,Walters1964,Walters1964a,Housiadas2012,Housiadas2016}, with the recent works revealing the coupling between force and rotation \cite{Castillo2019, Housiadas2019}. Similar forces and torques due to viscosity gradients have only recently been found \cite{Datt2019}
\begin{align}
    {\bf F}_{\eta} =& \, - 6\pi a \left( \eta_p \left( {\bf x}_c \right) - \eta_{p \infty} \right) {\bf U} + 2\pi a^3 \nabla \eta_p \times \bm{\Omega},\\
    {\bf L}_{\eta} =& \, - 8 \pi a^3 \left( \eta_p \left( {\bf x}_c \right) - \eta_{p \infty} \right) \bm{\Omega} - 2 \pi a^3 \nabla \eta_p \times {\bf U},
\end{align}
where $\eta_p \left( {\bf x}_c \right)$ is the polymer viscosity at center of the particle. The viscosity gradients introduce a coupling between force-rotation (or torque-translation) which is absent for spheres in homogeneous fluids. The additional force and torque in stiffness gradients are
\begin{align}
    {\bf F}_{\lambda} &= \frac{3\pi \eta_{p\infty}a}{20} \left( 2 (2 + \alpha) a^2 \bm{\Omega \Omega} - (7 + 5 \alpha){\bf UU}  \right) \cdot \nabla \lambda \nonumber\\
    &\,+ \frac{3\pi \eta_{p \infty} a }{40}  \left( -8(2 + \alpha) a^2  \left| \bm{\Omega}\right|^2 + \left( 3 - 5 \alpha \right) \left| \bf U \right|^2 \right) \nabla \lambda,\\
    {\bf L}_{\lambda} &= \frac{\pi \eta_{p \infty} a^3}{10}  (19 + 3 \alpha) \left( {\bf U} \cdot \bm{\Omega} \right) \nabla \lambda \nonumber\\
    &+ \frac{3\pi \eta_{p \infty} a^3}{20} \left( (1 + 2 \alpha) {\bf U} \bm{\Omega} - (39 + 8 \alpha) \bm{\Omega} {\bf U} \right) \cdot \nabla \lambda.
\end{align}
The stiffness gradients introduce an asymmetric coupling between force-rotation (or torque-translation): a particle that is translating but not rotating does not experience a torque in stiffness gradients, while a purely rotating particle can experience a force.

\textit{Active particles.}---We now consider active particles (${\bf u}^s\ne\boldsymbol{0}$). Biological swimming microorganisms are often neutrally buoyant, and so we assume here that the net buoyant force and torque acting on the active particles is zero. With no inertia and no external force, the hydrodynamic force or torque on the particle must be zero. Using Eq.~\eqref{eqn:velocities}, the translational velocity in an inhomogeneous Giesekus fluid is to the first order in $De$, $\varepsilon_{\lambda}$ and $\varepsilon_{\eta}$
\begin{align}
    {\bf U} = {\bf U}_N   & - \frac{2B_1B_2}{15a} \lambda_\infty \beta_\infty \left( 1 - \alpha \right) {\bf p}\nonumber\\
    &- \frac{a B_2 }{5}\beta_\infty\left( {\bf I} - 3 {\bf pp} \right) \cdot \nabla \left( \frac{\eta_p}{\eta_{p \infty}} \right),
    \label{eqn:translational}
\end{align}
where ${\bf U}_N  = U_N{\bf p}$. The correction to the velocity in homogeneous Newtonian fluid is simply a sum of corrections found previously in homogeneous viscoelastic fluids \cite{Datt2017, Datt2019a, Datt2020} (we'll denote this correction as ${\bf U}_{NN_\infty}$) and in a Newtonian fluid with viscosity gradients \cite{Datt2019}. The stiffness or relaxation time gradients do not affect the translational velocity to the first order. In both cases the corrections to the translational velocity are linear in the dipolar squirming mode, meaning the pullers are slower in viscoelastic fluids, and when swimming down viscosity gradients (while the opposite is true for pushers). Neutral swimmers on the other had swim at the same speed as in homogeneous Newtonian fluids.

Similarly, the rotation is to the leading order in $De$, $\varepsilon_{\lambda}$ and $\varepsilon_{\eta}$
\begin{multline}
    \bm{\Omega} = -\frac{1}{2}{\bf U}_N \times \beta_\infty\nabla \left( \frac{\eta_p}{\eta_{p\infty}} \right) \\ 
    + \frac{B_1B_2}{6a}\lambda_\infty\beta_\infty \left( \frac{2}{5} - \alpha \right) {\bf p} \times \nabla \left( \frac{\lambda}{\lambda_{\infty}}\right).
    \label{eqn:rotational}
\end{multline}
The rotation has the usual viscotaxis term proportional to $\nabla \eta_p$ \cite{Datt2019}. In viscosity gradients, pullers, pushers, and neutral swimmers all rotate to swim down the gradients or display negative viscotaxis. Additionally, the rotation has a new durotaxis term proportional to $\nabla \lambda$. Considering $\alpha = 0$ to eliminate any shear-thinning effects, we see that in stiffness gradients, pullers rotate to swim up the gradients or display positive durotaxis. Pushers do opposite by swimming down the gradients or display negative durotaxis. Neutral swimmers are unaffected by stiffness gradients to the leading order in $De$, $\varepsilon_{\lambda}$, and $\varepsilon_{\eta}$. These changes can be simply understood by noting that the polymeric stresses that cause pushers to speed up and pullers to slow down in homogeneous viscoelastic fluids \cite{Zhu2012}, are unbalanced (left/right) when there is a gradient in the elastic modulus of the fluid causing a rotation of the particle until it is aligned with the gradient; in fact, the correction in \eqref{eqn:rotational} can be simply written, using the results from \eqref{eqn:translational}, as 
\begin{align}
    \bm{\Omega} = -\frac{1}{2} {\bf U}_N  \times \beta_\infty\nabla \left( \frac{\eta_p}{\eta_{p\infty}} \right)  
    -c\frac{1}{2}{\bf U}_{NN_\infty} \times \nabla \left( \frac{\lambda}{\lambda_{\infty}}\right),
\end{align}
where the constant $c = (5/2)(2/5-\alpha)/(1-\alpha)$ simply depends on the shear-thinning parameter and $c=1$ when $\alpha=0$.

Overall, the particle rotation and the equilibrium orientation in an inhomogeneous viscoelastic fluid depends on the squirming ratio $B_2/B_1$ and the competition between durotaxis and viscotaxis which scales as $(B_2/B_1)De L_{\eta}/L_{\lambda}$ and if we assume these length scales and mode magnitudes are commensurate then it simply scales as $De$. In effect, durotaxis in viscoelastic fluids is significant at high Deborah numbers, a regime common for microorganisms \cite{Lauga2009}. Since both viscosity and stiffness gradients point in the same direction, pushers and neutral swimmers always rotate to swim down the gradients. Pullers on the other hand will swim up gradients if $(3/4)(2/5-\alpha)(B_2/B_1)(L_\eta/L_\lambda)De > 1$.

Durotaxis occurs over a timescale $\tau \sim \frac{20}{3\left|B_2/B_1\right|} \frac{1}{De \varepsilon_{\lambda}} \frac{a}{U_N}$. Assuming $\left| B_2/B_1 \right| \sim O\left( 1 \right)$ \cite{Gong2023}, and $De$ and $\varepsilon_{\lambda}$ to be as large as $0.1$ for our asymptotic analysis to hold, we get that $\tau \sim \left( 10^2 - 10^3 \right) \frac{a}{U_N}$. As particles can swim at $1-10$ times their body length per second, we get $\tau \sim 10-10^3\,\text{s}$. Durotaxis exceeds over rotary Brownian diffusion, $\tau_R^{-1} \equiv D_R = \frac{k_B T}{8 \pi \eta_s a^3}$, provided $\tau < \tau_R$, where $k_B$ is the Boltzmann constant and $T$ is the absolute temperature. Using properties of water, $\tau < \tau_R$ gives $a > 5\,\mu$m. Hence the impact of fluctuations decreases as particle size increases and becomes significant for particles smaller than $5\,\mu $m.

\acknowledgements
The authors gratefully acknowledge funding (RGPIN-2020-04850) from the Natural Sciences and Engineering Research Council of Canada (NSERC).

\bibliography{references}

\end{document}